\documentclass[universe,review,accept,moreauthors,pdftex]{Definitions/mdpi} 

\firstpage{1} 
\pubvolume{1}
\issuenum{1}
\articlenumber{0}
\pubyear{2021}
\copyrightyear{2020}
\externaleditor{Academic Editor: Francesco Shankar}
\datereceived{7 February 2022} 
\dateaccepted{16 March 2022 } 
\datepublished{} 
\hreflink{https://doi.org/} 

\def\be{\begin{equation}}
\def\ee{\end{equation}}
\def\ba{\begin{eqnarray}}
\def\ea{\end{eqnarray}}

\def\msun{M_\odot}

\def\ltsima{$\; \buildrel < \over \sim \;$}
\def\simlt{\lower.5ex\hbox{\ltsima}}
\def\gtsima{$\; \buildrel > \over \sim \;$}
\def\simgt{\lower.5ex\hbox{\gtsima}}
\def\ltgt{$\; \buildrel < \over \geq \;$}
\def\gtlt{\lower.5ex\hbox{\ltgt}}
\def\etal{{et al.}}

\Title{Dust in Clusters of Galaxies}

\TitleCitation{Dust in Clusters of Galaxies}

\Author{{Yuri A. Shchekinov} 
 $^{1,2,}$*$^{,\dagger}$\orcidA{}, {Biman B. Nath} $^{2,\dagger}$\orcidB{} and {Evgenii O. Vasiliev} $^{1,3,\dagger}$\orcidC{}}

\AuthorNames{Yuri Shchekinov, Biman. B. Nath, Egenii Vasiliev}

\AuthorCitation{Shchekinov, Y.A.; Nath, B.B.; Vasiliev, E.O.}

\address{%
$^{1}$ \quad {Lebedev Physical Institute of Russian Academy of Sciences, 53 Leninskiy Ave., Moscow 119991, Russia}; yus@asc.rssi.ru (Y.A.S.), eugstar@mail.ru (E.O.V.)\\ 

$^{2}$ \quad {Raman Research Institute, Sadashiva Nagar, Bangalore 560080, India}; {biman@rri.res.in}

$^{3}$ \quad {Institute of Astronomy, Russian Academy of Sciences, 48 Pyatnitskya St., Moscow, 119017, Russia}}

\corres{Correspondence: yus@asc.rssi.ru}

\firstnote{These authors contributed equally to this work.} 
\abstract{The presence of dust in the intracluster medium (ICM) has been a long-standing problem that is still awaiting elucidation. Direct observational diagnostics are rather challenging (though not impossible) either because of a sparse distribution of dust in the intracluster space that makes extinction measurements difficult or because of a low surface brightness of infrared emission from dust. Complex indirect approaches are currently available that can overcome uncertainties and provide a reasonable understanding of the basic regulations of the physical state of dust in the ICM. Contrary to the common opinion that the hot ICM does not allow dust to survive and manifest, many sparse observational data either directly point out that dust exists in the intracluster space or its presence is consistent with the data. Highly divergent data in direct evidence and highly uncertain indirect indicators are often connected either with dust fragility in a hot environment, the possible compactness of spatial (clumpy) dust distribution in the ICM, or dynamical features of dust transport. The source of dust is obviously connected with galaxies, and it turns out that in most cases, dust is carried from galaxies into the ICM while being thermally and dynamically shielded against the hostile influence of high-energy ions. In this review, we briefly discuss related issues from observational and theoretical points of view, including the transport of dust into the ICM, and the associated destructive and protective mechanisms and their characteristic time scales. }  


\begin{document}
\
\section{Introduction}
 
\subsection{Brief History} \label{hist}
Fritz Zwicky was the first to propose the existence of dust in galaxy clusters, considering the specific case of the Coma cluster \citep{Zwicky1951,Zwicky1962}. With our current knowledge of dust production by stars and dust optical properties, and {temporarily neglecting the dust-destroying property} of the hot and rather dense X-ray gas, we {can} roughly estimate {the} expected extinction in Coma. For the total mass of stars in the Coma $M_\ast\sim 10^{13}~M_\odot$,~a dust-to-star mass fraction of $\sim$$10^{-4}$, and a dust grain size $a\sim 0.03~\upmu$m, we can roughly obtain the  extinction $\tau_v\sim 0.3$, which is close to Zwicky's estimate $A_V\sim 0.4$. \citet{Karachentsev1968} confirmed this result with a similar estimate for Coma, with $A_V\sim 0.3$ and an averaged $A_V\sim 0.2$ for a sample of 15 clusters.  \citet{Bogart1973} and then \citet{Boyle1988} measured the reddening of galaxies in the clusters of the Abell group and inferred from their study the extinction in $B$-band $\sim$$0.2$. When these numbers are reduced to the dust-to-gas (DtG) mass ratio $\zeta_d$, the result is around 10\% of the Milky Way value $\zeta_{d,{\rm icm}}\sim 0.1\zeta_{d,{\rm MW}}\sim 6\times 10^{-3}$. Indirect arguments to support a  relatively high intracluster dust (ICD) content have been put forward by \citet{Romani1992}: they mentioned that optically selected quasars from~\citep{Veron1989} seem to avoid foreground Abell clusters with a considerable deficit: $\sim$$25$\% within $\sim$$1^\circ$ and up to $\sim$$65$\% within $20^\prime$--$40^\prime$ of the central cluster area. This finding allowed them to assume a sample average  extinction of $A_v\sim 0.4$ with  significant (up to $\Delta A_v\sim 1$) clumpiness.  

Rough estimates in the context of these findings indicate that dust particles do not suffer destruction in galaxy clusters. Indeed, if one assumes following \citep{Renzini1997} (see also references therein) the gas metallicity in the intracluster medium (ICM) to be $\sim$$0.3$ of the galactic value, one gets for a cluster of $R_{cl}\sim 3$ Mpc, and gas density $\sim$$10^{-3}$--$10^{-4}$ cm$^{-3}$, an optical depth of $\tau_v\sim 0.2$, provided that dust particles survive against destruction in the hot ICM. A  comparable estimate follows from very generic properties of X-ray gas and its metallicity in Coma $A_V\sim  0.1$, assuming the ICD to be insensitive to  destruction by a hostile ambient plasma and a similar dust-to-metal mass ratio to the mean value in the Milky Way ISM \citep{Polikarpova2017}. However, \citet{Maoz1995} argued that radio-selected quasars screened by foreground clusters  show nearly equal reddening compared to the unscreened one, and thus, the avoidance of optically selected quasars behind the clusters can be the result of observational selections. 

Several predictions of direct observations of infrared (IR) emission from intracluster dust have been attempted. \citet{Voshchinnikov1984} assumed that the amount of dust in the Coma ICM  corresponds to the extinction $A_v=0.3$ measured by \citet{Karachentsev1968} and obtained flux densities of $\sim$$3$ GJy sr$^{-1}$ in $\lambda=70$ to $200~\upmu$m waveband. Using a sample of 14 Abell clusters, \citet{Hu1985} have found $\sim$$10^5$ Jy sr$^{-1}$ based on their estimates of the X-ray gas mass in them. \citet{Dwek1990} calculated IR emission from dust in galaxy clusters, accounting for the stochastic heating of dust particles from hot ICM electrons, and estimated the flux density for the Coma cluster of $\sim$$0.1$ MJy sr$^{-1}$ at $\lambda=100~\upmu$m, which is almost two orders of magnitude lower than the {\it IRAS} value. However, when the total Coma flux is compared to the data of the {\it COBE} satellite, it results in a dust-to-gas mass ratio $\zeta_d=M_d/M_g=5.5\times 10^{-5}$, i.e.,  140 times depleted with respect to the Milky Way value, explicitly indicating that the ICD can suffer from the hostile influence of the hot ICM.    

Therefore, it is clear that the effects { of the very hot and rather dense intracluster plasma make it difficult  for a ``naked'' unshielded dust particle to survive on time scales comparable to characteristic crossing times for a typical cluster.} In particular, for the peripheral region of Coma ($R\geq 3$ Mpc), with $n\sim 2 \times 10^{-5}$ cm$^{-3}$, $T\sim 0.9$ keV ($\sim$$10^7$ K), the characteristic sputtering time for a particle with radius $\sim$$0.03~\upmu$m is $t_{sp}\sim 1$ Gyr, while the crossing time for Coma is $t_{cr}\sim 10$ Gyr. In the inner parts, say $R\simlt 1.5$ Mpc, the density is an order of magnitude higher, resulting in a proportional shortening of the lifetime. Therefore, from this point of view, dust particles can partly survive only in a circumgalactic coronae of radius $R_c\simlt 0.1$ Mpc, assuming the dust to be ejected out of a parent galaxy with $v_{out}\sim 100$ km s$^{-1}$. Smaller dust particles that usually determine extinction can occupy a  proportionally smaller volume around the parent galaxy. If, on the other hand, a fraction of smaller dust particles are hidden and shielded in denser and colder clouds of the ICM, they can contribute to a patchy extinction and reddening while {avoiding} IR emissions (see the discussion in \citep{Shchekinov2017}). Overall, it makes the problem of presence of dust in intracluster gas and correspondingly the detection of extinction or diffuse {IR} emission from ICM less obvious. This {puzzle} has stimulated much activity to search for any observational manifestation of dust ejected from parent galaxies into circumgalactic space and farther into the intergalactic space (see discussion in \citep{Nath1999,Ferrara1999,Bianchi2005,Montier2005,Menard2010,Shchekinov2011,Vasiliev2014,Zhu2014,Menard2015,Polikarpova2017,
Shchekinov2017,Erler2018,Melin2018,Vogelsberger2019,Gjergo2020} and references therein).  

\section{Later Observations} \label{latob}
However, later observations 
{ revealed the presence of dust in the diffuse gas of galaxy clusters with a much lower amount}. As an example, \citet{Stickel2002} have inferred from ISO observations at 120 and 180 $\upmu$m in five Abell clusters, including A1656 (Coma), a negligible dust extinction $A_v\ll 0.1$ mag, with a corresponding dust-to-gas mass ratio (DtG) $\sim$$10^{-6}$, { assuming a nearly Milky Way dust opacity. The inferred DtG is 55 times} less than that obtained by \citet{Dwek1990}. \citet{Chelouche2007} detected reddening $E(B-V)\sim (1-3)\times 10^{-3}$ toward several galaxy clusters with line-of-sights scanning distances up to $\simeq 7R_{200}$ from the cluster center ($R\leq R_{200}$ is the radius {within which the mean density} is 200 times of the {critical density}). The measured reddening shows flattening within the core $R<2.5R_{200}$, slowly decreasing as $\propto R^{-1}$ outside, which can be explained if the reddening is mostly determined by the contribution of peripheral cluster regions $R\simgt 3R_{200}$. {\rm The inferred DtG is around $\zeta_d<3.6\times 10^{-4}$, i.e., $<5$\% of the value in local interstellar medium (ISM). An important issue stressed by \citet{Chelouche2007} is that even though the observed reddening is definitely connected with dust, the data are insufficient for distinguishing the dust type. This is the main reason why most estimates of DtG in the ICM implicitly assume the Galactic value in local ISM.   } 

\begin{figure}[H]

\includegraphics[width=8.5 cm]{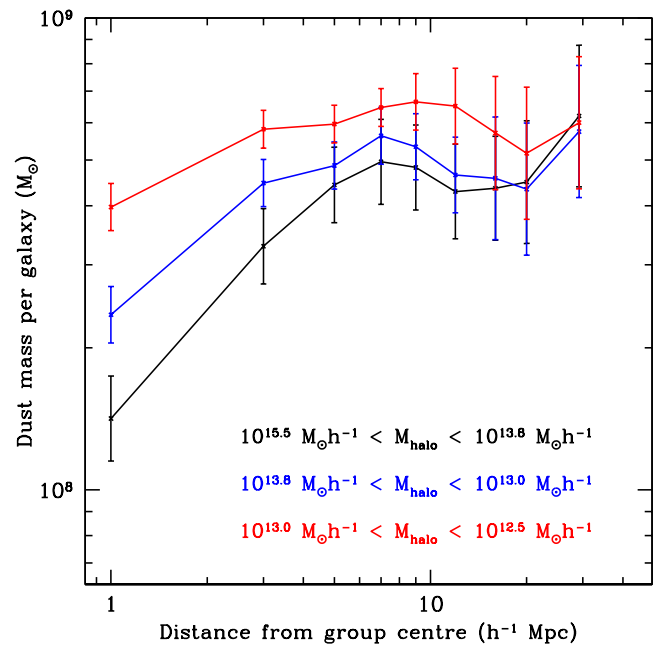}

\caption{The excess of dust mass per excess galaxy due to gravitational lensing versus the distance from the cluster (group) center. Three curves correspond to different mass bins of clusters from small to higher masses as follows from the uppermost to the lowermost; { see Figure 9 in \citep{McGee2010}. 
Reproduced with permission from S. L. McGee \&  M. L. Balogh, published by RAS (MNRAS), 2010.}}
\label{flatn}
\end{figure}

A similar phenomenon, that of a strong decrease of the dust mass per galaxy within the central 3 Mpc and its flattening outside as seen in Figure~\ref{flatn}, has been inferred by \citet{McGee2010} by examining the presence of dust in a sample of 70,000 of low-redshift groups and clusters in the Sloan Digital Sky Survey. {This can be interpreted as due to the fast destruction of dust in a hot  environment of central clusters' regions and partial survival in less hostile conditions at the  periphery.} In this connection, a very interesting and important result demonstrated in \citep{McGee2010} is that while lines of sight through a  massive cluster ($\sim$$10^{14}h^{-1}~\msun$) reveal a dust-to-gas mass ratio that is only 3\% of the local value in the Milky Way-- $\zeta_d\sim 2\times 10^{-4}$, in smaller galaxy groups ($\sim$$10^{12.7}h^{-1}~\msun$), this ratio is an order of magnitude higher: $\sim$$55$\%. { This result can be loosely interpreted in terms of a higher gas equilibrium temperature in more massive clusters, according to the `mass-temperature' relation $M_c\propto T^{\alpha}$ with $\alpha\approx 1.5$ see in \citep{Arnaud2005}.} 

\citet{Anand2022} have analyzed the cross-correlation of quasar MgII absorbers from the SDSS (Sloan Digital Sky Survey) DR14 with the galaxy clusters from the DESI (Dark Energy Spectroscopic Instrument) survey, and they concluded that the covering factor of cold absorbing clouds decreases from $\sim$$0.03$ in clusters centres to $\sim$$0.003$ in the outskirts. The equivalent width of MgII absorbers $W_\lambda\sim 0.3-1$\AA \ converges to the surface mass of MgII locked in the clouds is $\Sigma_{\rm MgII}\sim 1.3-4~\msun$ kpc$^{-2}$, which is equivalent, assuming the elemental abundance to be solar, to extinction $A_v\leq (2-6)\times 10^{-4}$, with the upper limit corresponding the ratio of dust to Mg mass fractions $\sim$$3$ as in the Milky Way. It is worth stressing that MgII absorbers are represented by small size dense and cold clumps, and this assumption of dust survival is fairly reasonable.  

More recently, \citet{Longobardi2020a} have found the presence of dust in the intracluster space of the Virgo cluster with $E(B-V)\simeq 0.042$ and extinction $A_v\simeq 0.14$, assuming the Small Magellanic Cloud extinction law. The estimated dust mass $M_d\simeq 2.5\times 10^9~M_\odot$, and the dust-to-gas mass ratio $M_d/M_g\simeq 3\times 10^{-4}$, i.e., nearly 20 times lower than in the Milky Way. The latter fact indicates that, as expected, dust in hot intracluster plasma suffers from destruction, mostly from thermal sputtering. 

A large scatter in the determined extinction $A_v\sim$ 0.01--0.1 or higher can be connected with observational uncertainties as discussed in \citep{Muller2008}, and with real variations optical properties of dust under hostile influence of hot ambient gas \citep{Shchekinov2011}.  In addition, the interrelation between the extinction and dust mass changes in the process of dust destruction: thermal sputtering destroys primarily dust grains of smaller size, whereas larger particles can survive longer, and as a result, the relation between the dust mass 
$ M_d\propto\int n(a)a^3da$ and its cumulative surface 
$S_d\propto\int n(a)a^2da$ 
changes. It can affect estimates of dust mass obtained in measurements of optical extinction and/or thermal dust emission. It follows from here that conclusions about dust optical properties and their relation to dust mass estimates can strongly depend on the efficiency of dust circulation in the mean field of intracluster space, i.e.,  the relation between the characteristic time of dust destruction in the ICM gas and dust replenishment from embedded galaxies.  

Direct observations of IR emission from hot intracluster dust also {demonstrate its rather diluted amount}. 
\citet{Bai2007} put forward only upper limits on IR emission at 24 and 70 $\upmu$m: $\sim$$5\times 10^3$ Jy sr$^{-1}$ and $\sim$$5\times 10^4$ Jy sr$^{-1}$, respectively, in the cluster Abell 2029. For the parameters of the Abell 2029 cluster, see e.g., in \citep{Walker2012}, this corresponds to an upper limit on dust content $\zeta_d\leq 2\times 10^{-7}$. Comparison of modeled IR spectra with the stacking IRAS measurements in $60~\upmu$m and $100~\upmu$m bands of a sample of $\approx ${13,800} 
 groups and clusters from the SDSS max-BCG catalogue resulted in a weaker upper limit on the intracluster dust mass fraction $\zeta_d\simlt 5\times 10^{-5}$ \citep{Roncarelli2010}. One of the difficulties related to observations of IR emission from the ICD is connected with the disentangling of contributions of the dust locked in galaxies and the dust between them \citep{Montier2005,Gutierrez2017}. This trouble has been stressed by \citet{Montier2005} after the detection of IR emission from a sample of {11,500} clusters in the IRAS data. The detected intensity was found to vary from $\sim$$3$ kJy sr$^{-1}$ at $12~\upmu$m to $\sim$$30$ kJy sr$^{-1}$ at $100~\upmu$m, which is consistent with the intensities expected from the intracluster dust. \citet{Gutierrez2017} have analyzed the FIR spectra of 327 clusters from the Herschel data to set a 95\% upper limit for the ICD of 86.6, 48.2, and 30.9 mJy per cluster at 250, 350, and 500~$\upmu$m, which translated to an upper limit of the dust-to-gas mass fraction $\zeta_d\sim 10^{-5}$, i.e., 0.03\% of the value in the Milky Way ISM. 

\subsection{Planck Measurements} 
\begin{figure}[H]

\includegraphics[width=9.5 cm]{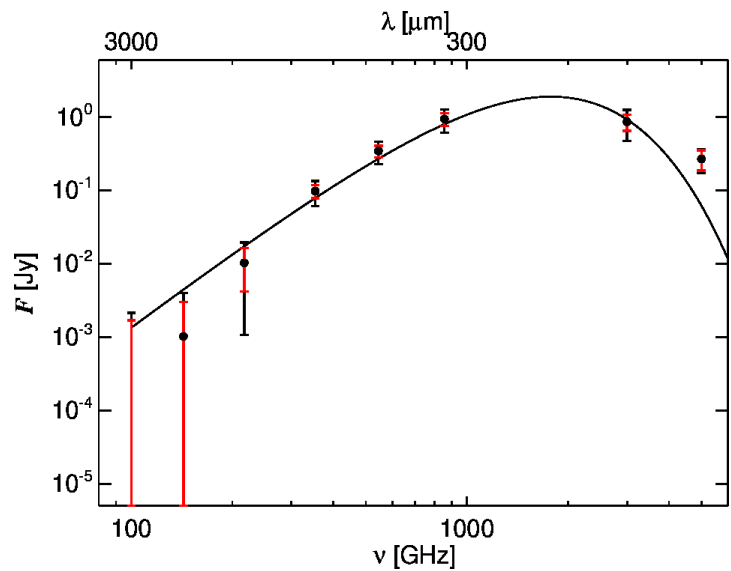}

\caption{{The cumulative} 
 spectrum averaged over the sample of 645 clusters with known redshifts $z\simeq$ 0.1--1 (black points); black error bars show the dispersion from a bootstrap resampling, while the red ones are obtained by integration at random positions 1$^\circ$ away from each cluster region. The solid black line shows the best-fit modified blackbody spectrum with the spectral index $\beta=1.5$. {Credit: \citet{Adam2016}.  A\&A, 596, A104, 2016, reproduced with permission \copyright~ ESO.}} 
\label{sed}
\end{figure}

Independent and complementary measurements of dust content in the ICM of the cluster sample with known redshifts in the range $z\simeq$ 0.1--1 via observations of thermal infrared dust emission have been obtained by {\it Planck} as reported in \citep{Adam2016}. It was their primary goal to separate dust emission from the one associated with spectral distortions due to the thermal Sunyaev--Zeldovich effect (tSZ). In the cumulative stacked signal, they found contributions from the tSZ in the submillimeter wave range and from the hot ($T_d\simeq 20$~K) dust at higher frequencies and with the stacked flux of $F_\nu\sim$ (0.001--0.01 MJy~sr$^{-1}${)}  
through over the intracluster space. The resulted spectral energy distribution (SED) is shown in Figure~\ref{sed}. \citet{Adam2016} stress that the obtained SED looks rather like the Galactic gray-body thermal dust emission $\kappa_\nu B_\nu(T)$ with $\kappa_\nu\propto\nu^\beta$, $\beta\approx 1.5$, the emissivity index. The two points, at 143 and 353 GHz, are seen to deviate downward and upward of the gray-body spectrum, and following \citet{Adam2016}, they represent a residual tSZ contribution. An important detail is that the tSZ and the hot dust emission have similar radial profiles in the clusters Figures 1 and 2 in \citep{Adam2016}, indicating that the dust indeed belongs to the diffuse intracluster space and is not confined in the cluster galaxies. A relatively high dust temperature, $T_d\simgt 20$ K, inferred by \citep{Adam2016} can be understood as a stochastic dust heating of small dust particles in hot plasma as demonstrated first by \citet{Dwek1990}.   

Given the fact that the presence of dust in clusters of galaxies (a) can have important implications on the interpretation of cosmological experiments, in particular, in spectral distortions of cosmic microwave background (see discussion in \citep{Adam2016,Melin2018,Zelko2021}), (b) can result in the incompleteness of galaxy clusters catalogue \citep{Melin2018}, and (c) can affect the thermodynamics of proto-cluster gas at $z\simgt 2$ when they are far from being virialized see, e.g.,  \citep{Casey2016,Cheng2019,Kubo2019,Smith2019a}, the question of whether and in what amount dust survives a long-term aggressive environment of intracluster plasma deserves detailed consideration.    

\section{Dust Circulation in ICM} \label{d_circul} 
\subsection{Dust Sputtering} 
As mentioned earlier, the ICM gas has a metallicity of $Z\sim 0.3Z_\odot$. Therefore, during the whole history of gas enrichment of the clusters' {inner region,} the total amount of injected dust can be as high as $M_d\sim 0.3ZM_g$, where $\zeta_d\sim 0.3$ is the mass-to-metal ratio commonly assumed in the Milky Way ISM, $M_g$ is the intracluster gas mass, which can be of \mbox{$M_g\sim 10^{12}$--$10^{13}~\msun$}, and as roughly estimated in Section \ref{hist}, dust extinction of \mbox{$A_v\sim$ 0.03--0.1} can be expected.       

However, what makes the question of dust presence in the ICM problematic (though not implausible) is the fragility of dust particles against the {destructive effect} of the hot plasma. One of the most efficient mechanisms of dust destruction is collisional sputtering---collisions of heavy high energy ions that remove atoms and ions from the surface of a target dust grain \citep{Draine1979,Jones1994,Dwek1996}. The threshold energy of this process is $E_{\rm sput}\sim 30$ eV and the sputtering yield is $Y(E)\sim 10^{-2}$ for protons to $Y(E)\sim 1$ for oxygen ions (see Figures 3 and 4 in \citep{Draine1979}). Therefore, the ICM {has suitable conditions} for dust sputtering, even though its density is low: from $n\sim 10^{-3}$ cm$^{-3}$ at the periphery $r\sim r_{500}$ to $n\sim 10^{-2}$ cm$^{-3}$ in the inner $r\simlt 0.1r_{500}$ regions, $r_{500}$ being the radius {within which the mean density is 500~times}  the background \citep{Ota2006,Ota2012}. With a simplified estimate of dust lifetime against sputtering $t_a(T)=a|da/dt|^{-1}\approx 10^{10} an^{-1}$ yr valid at $T>10^6$ K Eq. 25.14 in \citep{Draine2011}, one arrives at $t_{0.1~\mu{\rm m}}\sim 100$ Myr for dust grains of $a=0.1~\upmu$m at the peripheral cluster regions and \mbox{$\simlt$10 Myr} within $r<0.1r_{500}$, and proportionally shorter times for smaller grains. 

Therefore, when dust grains are ejected from galaxies into a direct contact with the hot ICM, they can occupy around the parent galaxy a layer of thickness $\Delta r\sim v_{ej}t_a(T)\leq v_{ej}t_{0.1~\mu{\rm m}}(T)$, where $t_{0.1~\mu{\rm m}}(T)$ is the lifetime of the grains with presumably the biggest radius $a=0.1~\upmu$m in the ISM. From the above estimates, the layer thickness can be \mbox{$\Delta r\leq 100v_{ej,100}n_{-3}^{-1}$ kpc}, $n_{-3}=10^3n$. Smaller particles are confined in thinner layers. The most numerous grains with $a=$0.01--0.03$~\upmu$m, which contribute predominantly to absorption and extinction, survive within only $\Delta r\leq (10-30)v_{ej,100}n_{-3}^{-1}$ kpc, though larger particles with $a\sim $0.03--0.1$~\upmu$m can occupy volumes of $\Delta r\sim$ 30--100 kpc. Therefore, the  naive understanding that dust cannot be extensively spread in the ICM is applicable only to dust of the smaller sizes.  
   
\subsection{Dust Injection from Galaxies} 

\subsubsection{Shock Driven Galactic Winds} \label{shockw}
One of the most important and widely discussed sources of dust supply into the intracluster space is thought to be connected with galactic winds driven by supernovae (SNe) activity during episodes of enhanced star formation---starbursts: \citet{Larson1974}, \mbox{\citet{Saito1979}}, \citet{Dekel1986,Suchkov1994,Maclow1999}, and \citet{Ferrara2000} argued that the energy emanated by SNe during such episodes can drive large-scale outflows and winds. Galaxies are known to spend a fraction of time in the phases of active star formation and starbursts: \citet{Hopkins2010} conclude that starbursts contribute around 10\% of the total star formation in local universe, \citet{Bergvall2016} {revised} this number to 4.4\%. However, more recently, \citet{Stuber2021} inferred $\sim$$20$\% fraction of galaxies with the central molecular outflows from a limited sample of nearby main sequence galaxies. Young massive stars formed during these phases inject huge energy into the ambient gas in the form of stellar winds or supernovae explosions via strong shock waves,  whose cumulative effect expels a considerable gas mass with galactic scale outflows; see the detailed review in \citep{Veilleux2020}. The expelled mass is in the form of cloudy fragments, as observed around galaxies with a strong wind. NGC 3079 is an example (see in \citep{Cecil2001,Hodges2020}), although in general, galactic winds are known to be multiphase by nature, { i.e., } they contain coexisting cold \mbox{(with $T\simlt 10^4$ K}) and hot ($T\simgt 10^6$ K) material \citep{Heckman1995,Martin2005,Heckman2009}. In these conditions, the survival time of dust in the ICM depends on the survival time scale  of the expelled gaseous fragments carrying the dust  while moving into the ICM. As stressed by \citet[][]{Ferrara2016}, dust is destroyed practically immediately, in $\sim$$10^4$ yr, after the disintegration of a shielding cloud in a hot wind environment. However, a common understanding is that the destruction time of clouds driven by ram pressure in the wind flow is shorter than the acceleration time~\citep{Klein1994,Cooper2009,Scannapieco2015,Schneider2017}; i.e., dense clouds are not entrained into the wind. This problem is known as the ``entrainment problem''; a detailed description is given in \citep{Zhang2017}. {The following two processes destroy dense clouds moving in the hot gas: hydrodynamical instabilities and thermal conduction. }

{\it {Hydrodynamic instabilities.} 
} A cloud of radius $R_c$ embedded in an ambient gas moving with  velocity $v$ is accelerated with characteristic time $t_{acc}\sim \chi R_c/v$, where $\chi$  is the density contrast between the cloud and the ambient gas $\chi=\rho_c/rho_a$. The cloud is subject to destruction mainly due to the shock wave propagating from the discontinuity between the cloud and ambient flow into the cloud center. It takes $t_{cc}\sim \chi^{1/2}R_c/v$, which is the time defined as the ``cloud crushing'' time \citep{Klein1994}. A similar order is obtained for the time scales of Kelvin-Helmholtz and Rayleigh-Taylor instabilities $t_{\rm KH}\sim t_{\rm RT}\sim\chi^{1/2}R_c/v$ for the most disruptive wavelengths $\lambda\sim R_c$ \citep{Klein1994}. Therefore, the acceleration time $t_{acc}$ is much longer, by a factor of $\sim$$\chi^{1/2}$, than the time when the cloud is destroyed $t_{cc}$. In numerical simulations, the disruption---mixing of $\sim$$50$\% of a cloud--- takes place on time scale $t\sim 4t_{cc}$ for $\chi=100$ to $\chi\sim 10t_{cc}$ for $\chi=3$ \citep{Klein1994}. Recent high-resolution simulations demonstrate that the scaled cloud destruction time $t/t_{cc}$ grows with the cloud radius and can become entrained into the bulk flow  \citep{Armillotta2017,Farber2021}.  

They also confirm that the clouds remain relatively cold ($T\simlt 10^4$ K) with densities in the range $n\simgt 0.02$ cm$^{-3}$ while being ejected out of the parent galaxy into the hot ($2\times 10^6$ K) low-density ($n=10^{-4}$ cm$^{-3}$) circumgalactic medium (CGM) \citep{Armillotta2017}, and as such, they can shield dust particles against hostile ICM environments. Such clouds with a large enough radius, $R_c\simgt 250$ pc, survive in a  typical CGM environment at least 250 Myr against hydrodynamic instabilities,  and with the initial ejection velocity of 100 km s$^{-1}$ can reach distances up to the circumgalactic region of 25 kpc with a relatively low level of mass loss---around half an initial mass at $t=250$ Myr, as demonstrated by \citet{Armillotta2017}. This trend, that of an enhancement of a cloud survival with its radius, remains valid for denser and colder clouds in the process of their acceleration by hot winds. A detailed consideration for clouds with the density contrast $\chi=10^3$ and temperature $T_c=10^3$ K immersed into a hot wind flow is given by \citet{Farber2021}. 

 It is worth stressing in this regard that direct convincing observations confirming the survival of dense (molecular) clouds ejected from galactic disks up to $\sim$$1$ kpc into the halo have been demonstrated to be present in the galaxy M55 \citep{Graham1982,Ferguson1996,Otte1999,Tullmann2003}. More recent observations confirm the existence of {in situ} ongoing star formation at $z\sim 1$ kpc in galactic halos~\citep{Stein2017,Howk2018}. Therefore, one can think that relatively dense and massive (presumably, molecular) clouds are elevated across such large distances and survive on time scales in the order of \mbox{$\simgt$1 Myr} until a considerable number of stars are born.   

{\it {Thermal conduction.}} When a dense and cold gas cloud is surrounded by a hot tenuous medium, the mutual action of thermal conduction and radiative cooling results in the formation of a thermal wave (TW) propagating from hot to cold phase, or the other way around, depending on conditions \citep{Balbus1985,Begelman1990,McKee1990,Ferrara1993a}. The velocity of the thermal wave is \citep{Ferrara1993a} 
\be \label{classic}
u_0\simeq {{\gamma-1}\over\gamma}\sqrt{\left({\kappa T\over n^2\Lambda}\right)}{\langle n^2\Lambda\rangle\over\rho_c c_h^2},
\ee
$\kappa$ [erg cm$^{-1}$ s$^{-1}$ K$^{-1}$] is thermal conductivity, subscripts $h$ and $c$ refer to hot and cold phases, $\Lambda$ is the cooling rate (averaging is carried out by the interface layer between the phases), $c_h$ is the sound speed in the hot phase. For the conditions of interests---a cloud immersed into a hot medium, in which $\kappa\approx 5\times 10^{-5}T^{5/2}$ is the classical Spitzer conductivity determined by hot electrons \citep{Spitzer1962}, one can write $u_0$ as \citep{Polikarpova2017} 
\be 
u_0\approx 1.8\times 10^6\chi^{-1}T_7^{3/4}~{\rm cm~s^{-1}},  
\ee       
$\chi=n_c/n_{\rm icm}$ is the density contrast between the ejected clouds and the ICM gas, \mbox{$T_7=T/10^7$}. {It follows that a cloud with radius $R_c$ can be evaporated in a hot medium within a time scale of $t_{ev}\approx 2\times 10^{12}\chi R_{c,\rm 1~pc}T_7^{-3/4}$ s, which gives $t_{ev}\sim 10-100 R_{c, \rm 1~pc}$ Myr for typical densities of ejected clouds $n_c\sim 0.01-0.1$ cm$^{-3}$ ($\chi=10^2$--$10^3$)}\endnote{\citet{Cecil2001} constrain the gas density in the ejected clumps and filaments $n>4.3f^{-1/2}$ cm$^{-3}$ with $f>3\times 10^{-3}$ being the volume filling factor of the clumps. \citet{Simcoe2006} have found a copious amount of relatively dense ($n_c\sim 10^{-3}-0.1$ cm$^{-3}$) clouds in the intergalactic medium at $z\simeq 2.28-2.29$ in the field of $500~h_{70}^{-1}$ (physical) kpc around the sight-line toward the QSO HS 1700+6416. The clouds in this field are metal enriched and presumably injected by galactic winds from galaxies in the same range of redshifts. Their densities are at least two orders of magnitude higher than in the ambient IGM at $z\simeq 2.3$.}. {It is seen that sufficiently dense clouds with sizes $R\simgt 100$ pc can survive thermal evaporation on Hubble time scales in the ICM with $T_7\simlt 1$.} The sizes of gas fragments and clouds close to the base of a galactic wind can vary from 1 to 100 pc as can be inferred from the data presented in \citep{Cecil2001,Bolatto2013,Salak2014}. When carried to the outer regions of the galactic halo, these clouds expand and considerably decrease in density depending on their velocities and initial thermal states as well as the ambient ICM pressure. However, as the density decreases as $n_c\propto R^{-1/3}$, the characteristic evaporation time will increase with cloud expansion. Therefore, one can expect that a fraction of the clouds injected from galaxies will eventually survive evaporative disruption within the Hubble time. Moreover, it is important to emphasize that the direction of the TW, i.e., evaporation or condensation, depends on the overall energy balance in the cloud--intercloud cooling--conductive interface. As shown first by \citet{Doroshkevich1981}, the TW in a cooling medium initiates condensation of the hot ambient gas and then the evaporation of an immersed cold dense cloud. 
A detailed discussion of this conclusion is given in \citep{Ferrara1993a,Meerson1996}. It is confirmed recently by numerical simulation that the contact of hot gas and a cold dense cloud in most cases drives the condensation TW that increases the amount of cold gas \citep{Kooij2021}.      

Another issue {worth mentioning} is connected with the fact that the free-path of electrons that transfer heat into cold clouds is large $\lambda_e\approx 10^4T^2/n_e$ cm \citep{Cowie1977}. The clouds of radius $R_c<0.3$ kpc ejected into a hot ($T\simgt 10^7$ K) and dilute ($n\simlt 10^{-3}$ cm$^{-3}$) ICM occur to be smaller than the free-path length of thermal electrons, and a consideration of their evaporation with a classic heat conductivity is not applicable for them. Such clouds undergo thermal heating with the saturated heat flux unless suppressed by a magnetic field~\citep{Cowie1977} 

\be 
q_{sat}\sim {3\over 2}k_{\rm B}nTv_{e,T}, 
\ee
with $v_{e,T}$ being the thermal velocity of electrons. The total heat gained to the cloud is balanced by the radiative cooling in its boundary layer of thickness $\Delta R_c$ determined by the condition 

\be 
\Delta R_c\simgt {3q_{sat}\over \langle\Lambda n^2\rangle},
\ee
with $q_{sat}$ being determined by the external hot gas, and radiative cooling in the denominator being determined by gas in the boundary layer. It gives an order-of-magnitude estimate $\Delta R_c\simgt 300 T_7^{3/2}n_c^{-2}$ pc, where  the assumed ICM density  $n\sim 10^{-3}$ cm$^{-3}$, the cooling function in the boundary layer is assumed tobe $\Lambda\sim 10^{-23}$ erg cm$^3$ s$^{-1}$, which is valid for a non-steady state cooling gas \citep{Vasiliev2013}. Relatively dense clouds close to the parent galaxies with $n_c\sim 0.3$ cm$^{-3}$ and $R_c\simgt 1$ kpc can protect dust particles of destruction. However, larger clouds with lower density that enter deeper into the ICM can suffer from thermal destruction along with the dust inside unless thermal waves propagate in the opposite direction---from cold clouds to hot ICM, as discussed above. From a theoretical point of view, the role of thermal waves in cloud disintegration in the ICM still remains to be studied in detail. Hydrodynamical instabilities in the process of relative motions of clouds into ambient ICM medium appear to be the prevailing sources of clouds, and consequently, dust destruction.    

From an observational point of view, the sensitivity of the dynamics of clouds ejected from galaxies into the ambient extragalactic space, their survival, and their ability to transfer enriched material out of galaxies can be illustrated by two examples of galaxies: M82 and NGC 4945, where the galactic wind has been scrutinized with high angular resolution at a ``microscopic'' level. 
  
\begin{figure}[H]

\includegraphics[width=9.5 cm]{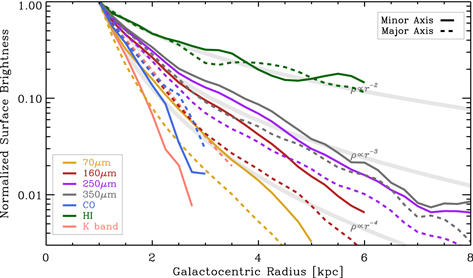} 

\caption{The radial profiles of gas tracers along the major and minor axes. The intensities are normalized to the one at $r=1$; the $x$-axis shows a distance from the center of M82 along the radial (major axis) or vertical (minor axis) directions. Shown are: dust emission at $70,~160,~250, and~350~\upmu$m, $^{12}$CO($J=2-1$) 1.3 mm, HI and stellar emission in $K$-band 2.159 $\upmu$m, as indicated in legends. Credit: \citet{Leroy2015}. Profiles along the minor and major axis are depicted by solid and dotted lines. Radial profiles $r^{-2}$, $r^{-3}$ and $r^{-4}$ shown in gray illustrate diverging and converging constituents when { integrated from \citep{Leroy2015}, their Fig. 13. Reproduced with permission from A. K. Leroy \etal, published by AAS (ApJ) 2015.}} 
\label{m82emis}
\end{figure}

{\it {M82 case.}} A good illustration of high-resolution data is given by \citet{Leroy2015} from a detailed analysis of  HI, CO(2--1) and dust emissions from outflowing gas in the galaxy M82. They conclude that the cold gas outflow  at a height $z\sim$ 1--2 kpc along the minor axis widens its collimation angle and exhausts, whereas warmer components, including H$\alpha$, HI, and PAH emissions within the hot wind cavity lift {above} $z\simgt 3$ kpc. Basic cold gas tracers are seen in Figure~\ref{m82emis}. HI is seen to extend along the outflow core (minor axis) to large distances with a shallow profile of $\sim$$r^{-2}$, indicating that the HI mass flux ``diverges'' in the sense that it does not decline at the heights comparable to the disk radial extent. On the other hand, all other profiles belonging to cold components lie below the $r^{-2}$ line and thus show ``converging'' fluxes within relatively narrow heights $z\sim 2$ kpc.     

{\it {NGC 4945.}} However, an opposite example 
 is given by \citet{Bolatto2021} for the galaxy NGC 4945, which is another nearby starburst galaxy. From an analysis of the CO($J=3-2$) line with $\sim$$0.3^{\prime\prime}$ ($\sim$3 pc) resolution, the molecular outflow in the form of plumes shows velocities up to $\simgt 600$ km s$^{-1}$, which is sufficient to escape the galaxy's gravitational potential and reach its circumgalactic halo, as stressed by \citet{Bolatto2021}. Note that apart from CO lines, the associated lines of HCN and HCO$^+$ molecules, known as traces of high-density gas, are also detected, indicating that outflowing gas may have densities of $10^5$--$10^7$ cm$^{-3}$. Then, one can expect that clumps and clouds with such a high density will survive disruption by hydrodynamic instabilities in their trajectory into the CGM.
 
\begin{figure}[H]

\includegraphics[width=9.0 cm]{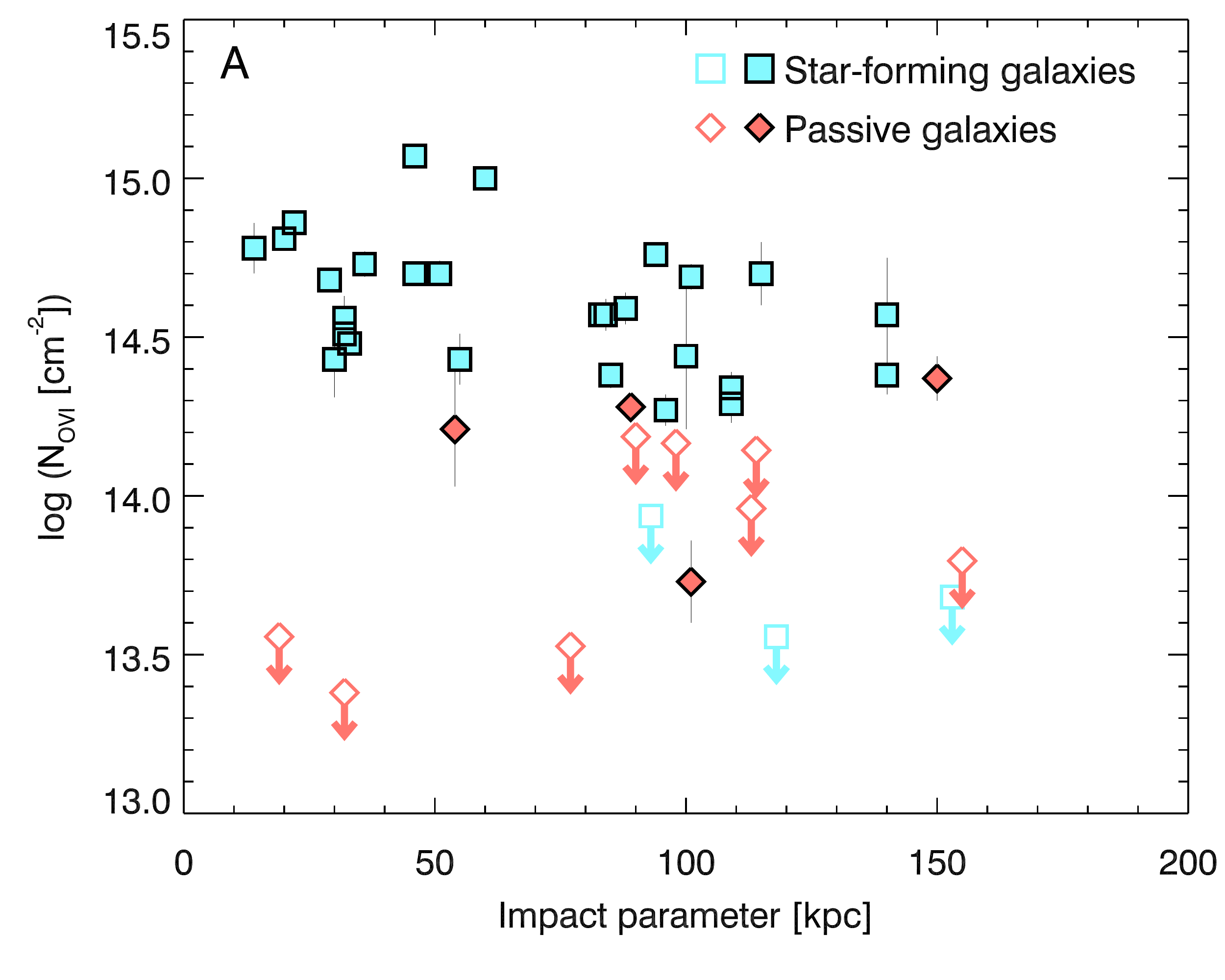} \\
\includegraphics[width=9.5 cm]{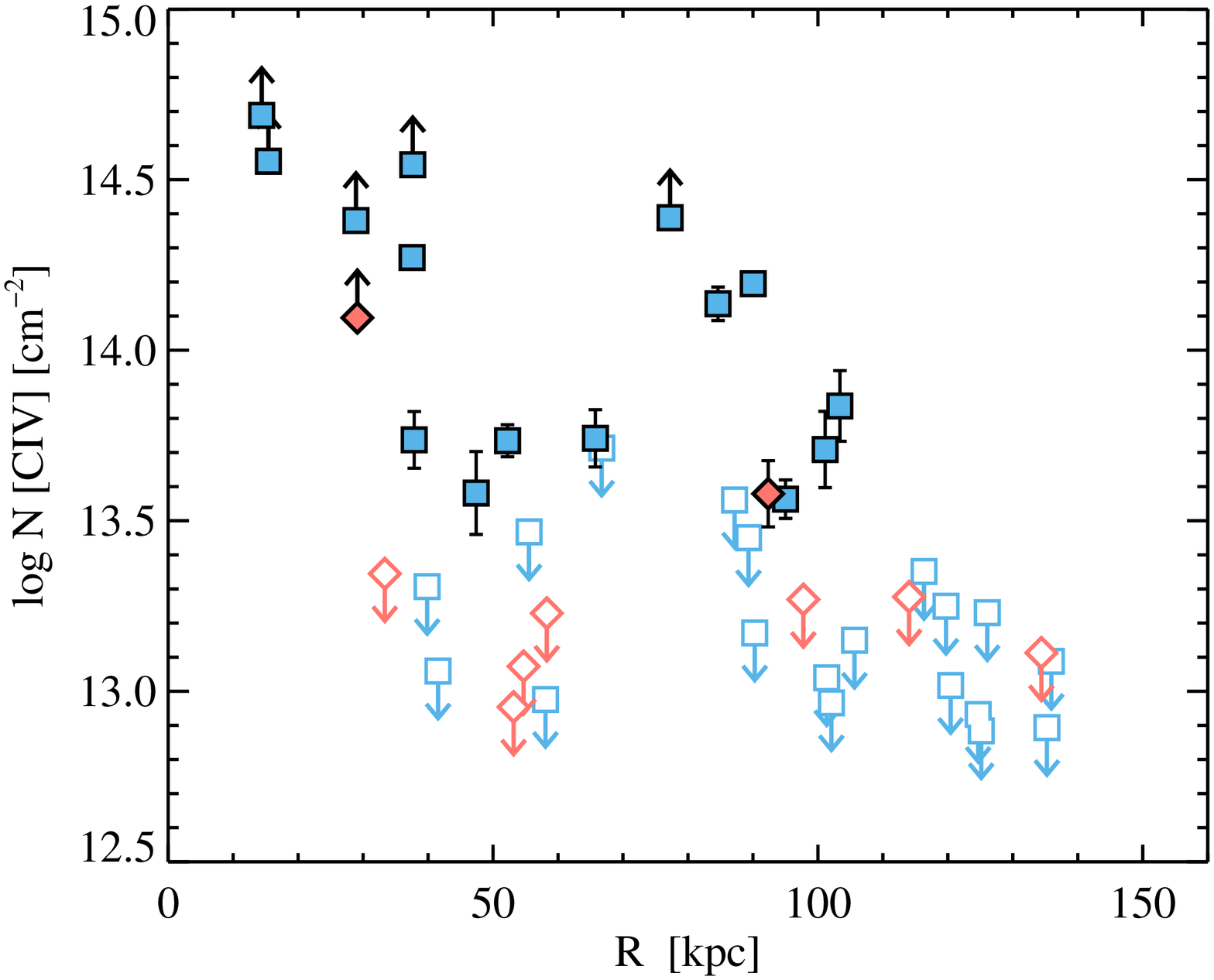} 

\caption{The radial profiles (in terms of the projected radius) of OVI ions in circumgalactic halos of the COS-halos survey---({\it upper panel}). Credit \citet{Tumlinson2011}, and of CIV ions in CGM of the COS-dwarfs survey ({\it lower panel}) { from \citep{Bordoloi2014}, lower left panel on their Fig. 2. Reproduced with permission from R. Bordoloi \etal, published by AAS (ApJ) 2014.} Blue squares and red diamonds represent column densities in star-forming and passive galaxies, correspondingly; the filled symbols depict detections, while the open with downward arrors depict non-detections within $2\sigma$.  } 
\label{cgmion}
\end{figure}

In spite of these dynamical details of mass transfer from galaxies into external medium at a ``microscopic'' level, which are not yet well understood, {the fact that the mass transfer is rather efficient} is illustrated by the existence of large---150 to 200 kpc, highly metal-enriched circumgalactic halos amounting to total amount of metal mass comparable to that in the disk of their host galaxies\endnote{Obtained with the Galaxy survey with the Cosmic Origin Spectrograph on the Hubble Space Telescope.} \citep{Tumlinson2011,Bordoloi2014,Tumlinson2017}. The radial distribution of column densities of OVI and CIV ions from these surveys is shown in Figure~\ref{cgmion}. Such halos are intermediate stages in the overall transport of metal-enriched material from galaxies into IGM. They also confirm that clouds that carry metals from galaxies are eventually destroyed when they {are fairly deep inside} the CGM.  

Turning to the ICM, the very existence of such extended and metal-enriched circumgalactic halos indicates that ejected dense clumps and clouds lose their mass on scales comparable to the sizes of the circumgalactic (CG) halos---150 to 200 pc. As seen from the estimates of cloud disruption by thermal conduction that have been stressed by \citet{Armillotta2017}, larger clouds lose their mass on longer distances and are less sensitive to the initial ejection velocity. This can mean that in reality, the mass spectrum of ejected clouds covers a wide range including those able to survive disruption while traveling galacto-centric distances of 150--200 kpc. This assumption is implicitly reflected in the fact that the ICM gas metallicity is close to 0.3 of the solar value $Z_\odot$ \citep{Renzini1997}. Given the fact that the ICM gas mass considerably exceeds the mass of galaxies, its high metallicity implies an efficient exchange of metal-enriched mass between the galaxies and the ICM. In field galaxies, the characteristic size of the enriched CG halos 150--200 kpc represents the scales of the dust survival region. However, in galaxy clusters, more frequent tidal interactions and a stronger ram-pressure stripping effect from the ICM likely limit these scales; see for more details the recent discussion in \citep{Boselli2021,Molnar2021,Muller2021}. 

\subsubsection{Radiatively Driven Galactic Winds}
Another important mechanism of expelling dust from galaxies is connected with radiation pressure on dust particles. A series of studies has been focused on the vertical support and transport of dust particles in optically thin low-density galactic halos. The amount of  halo gas above 1--2 scale heights is only $\leq e^{-1}$ of the total column  density, and correspondingly, radiation-accelerated dust particles move without transferring their momentum to gas, i.e., without a launching of galactic bulk mass outflows  \citep{Ferrara1990,Franco1991,Ferrara1991,Ferrara1993b,Shustov1995,Bianchi2005,Vasiliev2014,Sivkova2021}. Indirect considerations of radiation pressure as a  driving mechanism of dust acceleration in the form of filaments and their substructures extending up to $\simgt 1$ kpc above the plane of the NGC 891 galaxy have been obtained by \citet{Rossa2004}. A noteworthy conclusion of \citet{Rossa2004} is that dust and H$\alpha$ filaments do not correlate spatially, which confirms that the former is formed due to radiation pressure. Similar conclusions also follow for extra-planar dust filaments in the NGC 4217 galaxy as was argued by \mbox{\citet{Thompson2004}}. 

PAH molecules found in the NGC 5907 and NGC 5529 galaxies \citep{Irwin2006,Irwin2007} can also provide convincing signatures of radiatively driven dust, because shock waves efficiently destroy dust. Later on,  three edge-on galaxies---NGC 891, NGC 5775, and NGC 4044---have been confirmed to reveal filamentary structures in 8$~\upmu$m PAH feature extending up to $\sim$1--2 kpc \citep{Rand2011}. More recently, \citet{Sethi2017}, based on observations of an absorption feature at $\lambda\approx 6565$ \AA\, through over a large area on sky by \citet{Zaritsky2017}, have  hypothesized that PAH molecules can widely pervade the Milky Way gaseous halo. 

It is important to emphasize that disentangling the actual source of momentum for driving dust out of  galactic plane---large-scale shock waves versus radiation pressure--- is always challenging. However, as pointed out by several authors, stellar radiation can be efficient in galaxies only above $\sim$$1-2$ scale heights of the interstellar gas, whereas at lower heights, other mechanisms---SNe shocks and/or convection---are to be the dominant carriers of dust from the plane upwards \citep{Ferrara1993b,Shustov1995,Schroer2000,Bianchi2005,Dettmar2005,Sofue1994,Vasiliev2014,Sivkova2021}. This is connected with the fact that at heights $z\simlt 1-2$ of scale heights with a relatively high gas density, collisional coupling with the gas and a corresponding collisional drag force are large, such that the momentum from stellar radiation is insufficient. On the other hand, local outflows and bubbles driven by SNe shocks from small star clusters $M\sim 10^3-10^4~\msun$ extend within 1--2 scale heights and produce in the ISM rather regular vertical structures, such as Galactic `worms' by \citet{Heiles1992} and those identified by \citet{Sofue1994} in the NGC 253 ``boiling'' galaxy. Then, stellar radiation pressure facilitates  the separation of dust from gas and elevates dust to higher heights $z\sim 1$ kpc, where it is observed in the form of rather irregular filaments \citep{Rossa2004,Thompson2004,Rand2011}.   

Later on, it was pointed out by many authors that the momentum supplied into the ISM by stellar radiation can play an important dynamical role \citep{Scoville2003,Martin2005,Murray2005,Thompson2005,Krumholz2009,Murray2011,Gupta2016}. \mbox{\citet{Murray2005}} paid attention to the fact that radiation \textls[-5]{pressure from a starburst is comparable to the ram pressure from shock waves by SNe produced in the starburst. \citet{Martin2005}, \mbox{\citet{Murray2005}},} and \citet{Nath2009} confirmed that the wind regime with the radiation pressure as a driver is in  general agreement with observational correlations in ULIRGs. In particular, the growth of maximum velocity of cold clouds entrained in the outflow with the SFR and a Faber--Jackson-like  correlation between the SFR and the dispersion velocity $L\propto\sigma^4$ for a set of galaxies with bright energy sources in their central regions. \mbox{\citet{Hopkins2012}} have demonstrated that radiative-driven winds provide the mass-loading factor $\dot M_{wind}/{\rm SFR}\propto v_c^{-1}$ with $v_c$ being the galaxy circular velocity, and as a result, they dominate in galaxies more massive than the Milky Way---$M_g\simgt M_{\rm MW}$. It is also noteworthy that radiation pressure-driven winds in disk galaxies typically have rather high terminal velocities considerably exceeding their escape velocities $v_\infty\sim (2-3)v_{esc}$ \citep{Sharma2011}, and as such, they can be beneficial for supporting efficient galactic winds. 

Recent numerical experiments demonstrated that radiation pressure is more efficient in the acceleration and entrainment of optically thin gas clouds, and it does not initiate hydrodynamic instabilities \citep{Murray2011,Zhang2018,Huang2020}. In this context, IR radiation from energy sources can be more efficient in the acceleration of clouds than optical or UV radiation. \citet{Zhang2018b} have concluded that clouds being accelerated by a sufficiently strong IR radiation flux survive longer.           

\subsubsection{Cosmic Ray-Driven Galactic Winds}
The dynamical role of cosmic rays (CR) in galaxies is comparable with that of thermal energy, the energy of turbulent motions, and the energy of magnetic field; see for a review~\citep{Zweibel1997,Beck2001,Beck2012}. Therefore, one expects that cosmic rays can also drive galactic outflows and transfer gas and dust into circumgalactic and intergalactic medium. \citet{Ipavich1975} was probably the first who paid attention to the ability of CR to initiate large-scale motions in the ISM. Further on, this possibility has been considered in \citep{Breitschwerdt1991,Ptuskin1997} and more recently developed in \citep{Samui2010,Uhlig2012,Booth2013,Farber2018,Fujita2018,Jana2020,Bustard2021} and references therein. 

The two distinct features of CR-driven winds, that they are highly efficient mass loading and have a less destructive action on multiphase media, make them promising transport mechanisms for supplying dust particles into the intergalactic and intracluster space. \citet{Samui2010} and \citet{Booth2013} have shown that galactic winds maintained by CR can involve into the outflow up to a $\eta_w\sim$ 0.2--0.5 fraction of the SFR for Milky Way-class galaxies, and scale as $\propto v_c^{-2}$, so that  for less massive galaxies $\sim$$10^8-10^9~\msun$, it may be $\eta_w\simgt 10$, whereas it is only $\eta_w\sim 1$ without CR \citep{Booth2013}. On the contrary, CR-driven winds for very massive galaxies---$M\sim 5\times 10^{12}~\msun$---are shown to have only $\eta_w\simlt$ 0.006--0.03~\citep{Fujita2018} and drop with the galaxy mass as $\eta_w\propto M^{-\alpha}$, $\alpha\sim 1-2$ \citep{Jacob2018}, as if CR sink in massive galaxies. 
A detailed comparative analysis of different models of CR-driven galactic winds can be found in \citep{Jana2020}. 

Recent numerical simulations show that CR wind is less aggressive to clouds and accelerates them more efficiently  \citep{Wiener2019,Scannapieco2020,Bustard2021}. A drop in the Alfv\'en speed at the boundaries of dense clouds results in a build up of the CR pressure in the frontside of a cloud edge and ensures its effective acceleration. On the other hand, CR streaming along the magnetic field on the  cloud boundaries form the shear layer that is unstable against KH instability. Eventually, a cloud being accelerated by CR-driven wind gets destroyed and loses 50\% of its mass, on a time scale $\approx 12t_{cc}$, i.e., three times longer than in the case of destruction by a shock wave. A decrease in the cloud density contrast $\chi$ and the plasma $\beta=P_{th}/P_B$ prolongs the destruction time scale \citep{Scannapieco2020} due to diminishing Alfv\'en speed and restriction of the growing CR pressure at the frontside cloud boundary.  

\subsubsection{Tidal and Ram-Pressure Stripping} 
A widely discussed mechanisms of dust supply from cluster galaxies into the intracluster space is connected with tidal interactions and the dissipation of galaxies, which is revealed by the so-called intracluster light (ICL) first detected by \citet{Zwicky1951} and confirmed later by \citet{Matthews1964}. The ICL is presumably produced by a population of stars permeating throughout the cluster and gravitationally bound by its potential; see the discussion in \citep{Gonzalez2005,Dolag2010}, and references therein. Up to 35\% of the total stellar population in clusters can be {part of this population of} intracluster stars \citep{Gonzalez2007}. 

\begin{figure}[H]

\hspace{0.3cm}
\includegraphics[width=7.5 cm]{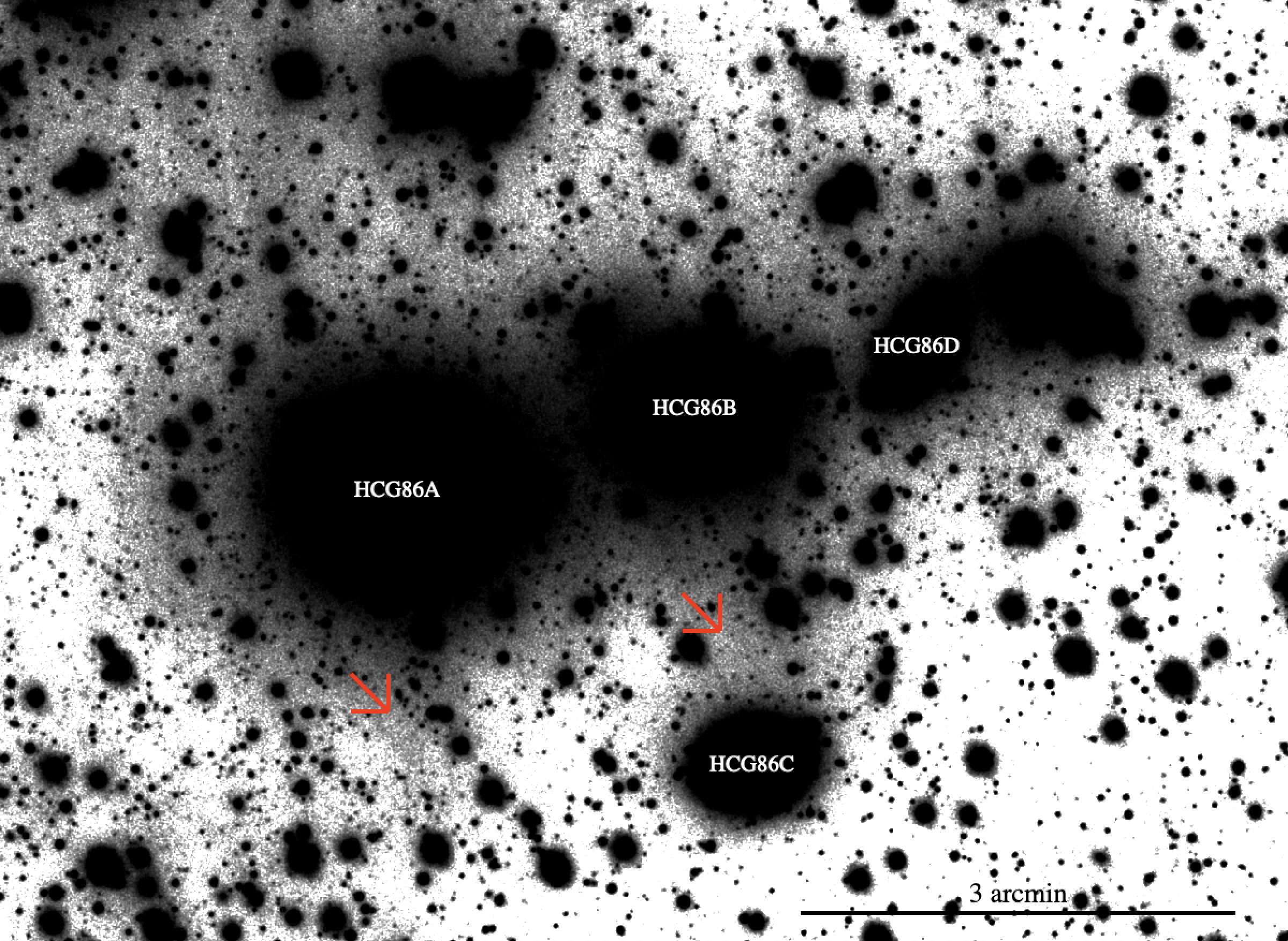} \\
\vspace{0.5cm}
\includegraphics[width=8.5 cm]{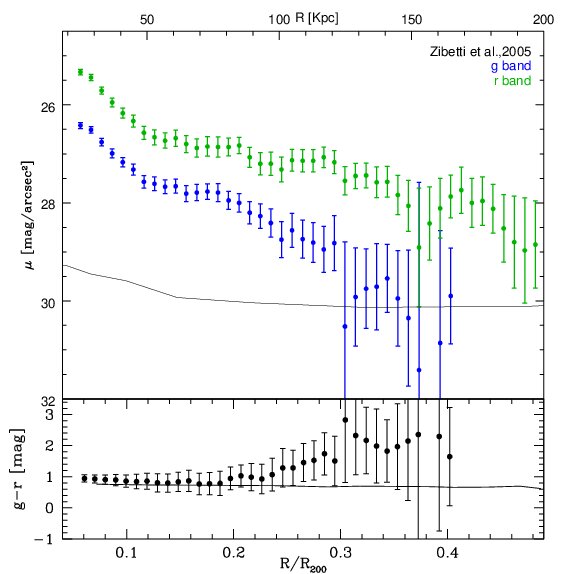}

\caption{{\it Upper panel:} The central part of the cluster group HCG 86, with the four dark spots being the brightest group members. Low-surface brightness filaments and bridges represent the ICL (two of them are indicated by red arrows). The {\it lower panel} shows azimuthally averaged color ({\it g} and {\it r}) radial profiles (points with error bars) and compares them with the stacked radial profile (solid line) for a set of $\sim$$600$ galaxy clusters from \citep{Zibetti2004}; HCG 86 and clusters' radii normalized to their $R_{200}$. { Credit:  \citet{Ragusa2021}.  A\&A, 651, A39, 2021, reproduced with permission \copyright~ ESO.} } 
\label{hcg}
\end{figure}

Galaxy groups also reveal a considerable fraction of their light being spread diffusely in the intergroup space. A good example has been recently reported in observations of the compact group of galaxies HCG 86 by \citet[][]{Ragusa2021}. The group demonstrates the presence of a substantial amount of low-surface brightness ICL collected in fragmentary structures and bridges between interacting galaxies. Their sizes can be of $\sim$10--30 kpc at a distance of HCG 86 $d\simeq 82$ Mpc, and they can pervade the intragroup space (see upper panel on Figure~\ref{hcg}). With the velocity dispersion of group members of \mbox{$\sim$$350$ km s$^{-1}$}, the characteristic lifetime of the filamentary features can be of the order of 60 Myr, assuming them to be temporary remnants of intergalactic tides. The azimuthal average surface brightness and the color of these fragmentary features in {\it r} and {\it g} bands at \mbox{$r\simgt 100$ kpc}, as seen in the {\it lower panel} on Figure~\ref{hcg}, extend up to $r\sim 0.5R_{200}$,  { over and above the contaminations}  from Galactic cirrus at $\mu\sim 30$ mag arcsec$^{-2}$. If this emission stems from stars, it would correspond to stellar density of $\sim$$0.1-1~\msun$ pc$^{-2}$, suggesting that they belong to galactic outskirts. However, \citet{Ragusa2021} conclude that the ICL colors $g-r\sim 0.7$ mag and $r-i\sim 0.4$ mag are consistent with the stellar population synthesis of intermediate-mass ($M=10^{10}-10^{11}~\msun$) galaxies without clear traces of dust, besides contaminations from Galactic cirrus at $\lambda=100~\upmu$m. It is worth noting that the averaged radial profiles in HCG 86 look similar (although obviously brighter) to the ones inferred for galaxy clusters \citep{Zibetti2004,Zhang2019}, with the floor at $\mu\sim 30$ mag arcsec$^{-2}$ strongly contaminated by the Galaxy cirrus \citep{Mihos2017,Kluge2020,Li2021}.    

\begin{figure}[H]

\includegraphics[width=12.0 cm]{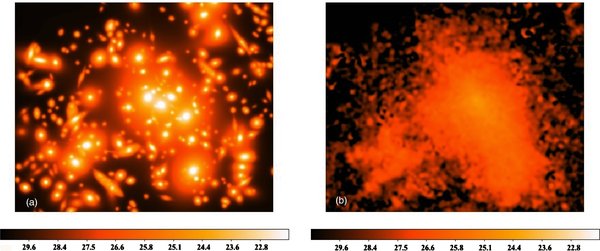} \\
\vspace{0.5cm}
\includegraphics[width=8.0 cm]{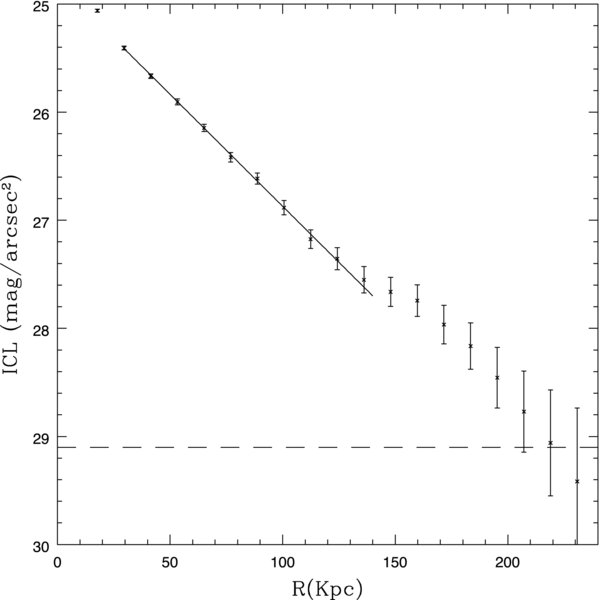}

\caption{The LBT image of the central ($R\sim 200$ kpc) galaxy cluster CL0024+17, $z\sim 0.4$, the field size is $95\times 77$ arcsec$^2$. The {\it upper left panel} shows only galaxies and diffuse light around them, on the {\it upper right panel} only residual ICL is shown, the {\it lower panel} confirms this conclusion: the inferred ICL radial profile is centered at its maximum and extends up to $\simgt 200$ kpc, error bars indicate uncertainties connected partly with background { subtraction from \citep{Giallongo2014}, their Fig. 6 and 7. Reproduced with permission from E. Giallongo \etal, published by AAS (ApJ) 2014.}}
\label{iclvir}
\end{figure}

A clear observational evidence of the virialization can be recognized in the image of the cluster CL0024+17 on Figure~\ref{iclvir} obtained at the Large Binocular Telescope (LBT) by \citet{Giallongo2014}: in the {\it lower panel}, a concentration of the ICL toward the apparent cluster center of mass is observable. It is seen that the ICL occupies mostly the central cluster region with radius of $r_{icl}\simgt 200$ kpc, although a considerable part of it can be hidden beneath the background. 

It is  reasonable to assume that the tidally stripped interstellar gas and dust can be connected with the ICL, because tidal forces equally act on  stars and  all the constituents of interstellar gas: dust, magnetic field, and clouds, as demonstrated first by \citet{Toomre1972,Wright1972}, and widely discussed recently in \citep{Teyssier2010,Renaud2015,Elmegreen2016,Smith2019b,Lizee2021}. Direct observational evidence was recently presented in \citep{Longobardi2020a,Longobardi2020b}. Intracluster light is commonly observed in galaxy clusters see, e.g., \citep{Feldmeier2004,Mihos2004,Gerhard2005,Mihos2005,Krick2007,Krick2011}, and references therein, and typically shows central concentration \citep{Krick2007,Dolag2010,Krick2011,Giallongo2014}, indicating that the tidal interaction {contributes to} the process of virialization. For a recent discussion, see  \citep{Ko2018,Montes2019,Contini2021} and the reviews by \citet{Mihos2016} and \mbox{\citet{Contini2021b}}. 

\begin{figure}[H]

\includegraphics[width=9.0cm]{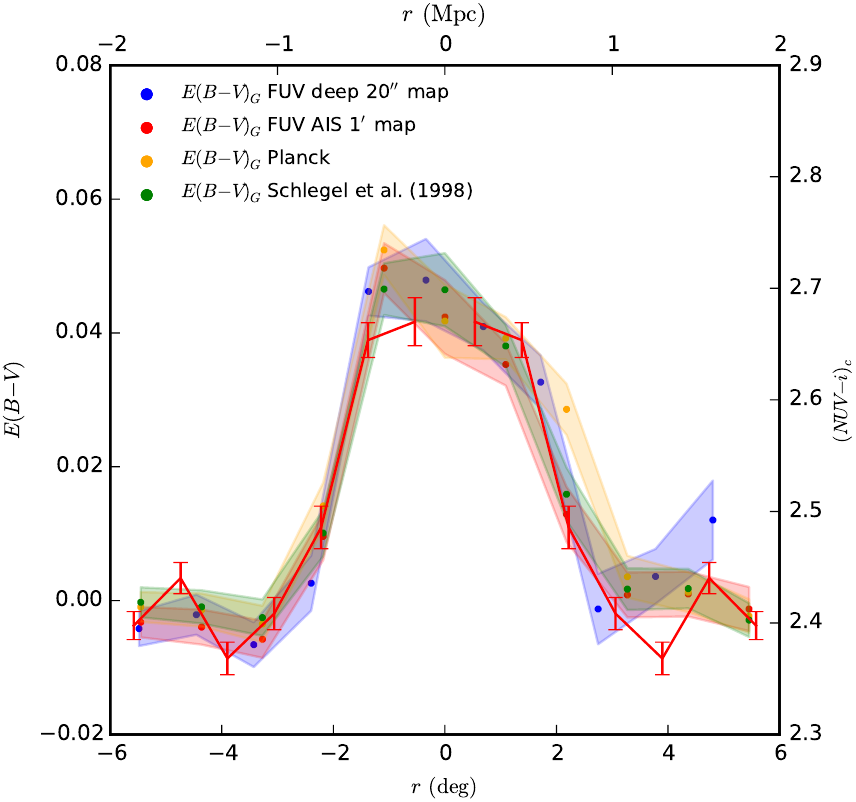}

\caption{
Radial $E(B-V)$ reddening profile (filled circles + uncertainties as shaded area), right $y$-axis show $(NUV-i)$ color, 5$^\circ=$1.44 Mpc, from negative to positive is radial direction from the southern to northern parts of the Virgo cluster, the red line with error bars shows colors averaged over the north and south parts of the cluster; different colors of circles correspond to corrections for galactic contaminations, making use of different galactic FUV  maps as described in { \citep{Schlegel1998,Planck2014,Boissier2015}, green, yellow, red 1$^\prime$, and blue 20$^{\prime\prime}$, correspondingly. Credit: \citet{Longobardi2020a}.  A\&A, 633, L7, 2020, reproduced with permission \copyright~ ESO.}}   
\label{dustvirgocolor}
\end{figure}

One may surmise that the central regions of galaxy clusters with frequent tidal interaction and gravitational relaxation to a virial state can contain in the ICM some tracers of interstellar gas and dust lost from galaxies acting in mutual tides. However, accounting for the fact that the cluster's central regions $R\simlt 100$ kpc are rather dense ($n\simgt 10^{-2}$ cm$^{-3}$ and hot $T\simgt 10^7$ K), dust particles do not survive longer than 10 Myr. It is also difficult to envisage that dense and cold clouds that shield dust particles can survive against a violent mixing in the process of virialization in cluster centers on long time scales. Nevertheless, \citet{Longobardi2020a} present an $E(B-V)$ radial profile for the Virgo cluster that clearly peaks at its center (Figure~\ref{dustvirgocolor}). A speculative consequence of this fact can be that the rate of tidally stripped gas and dust in the central Virgo cluster is sufficiently high to replenish {quickly destroyed dust}. 

\unskip

\begin{figure}[H]

\includegraphics[width=9.0cm]{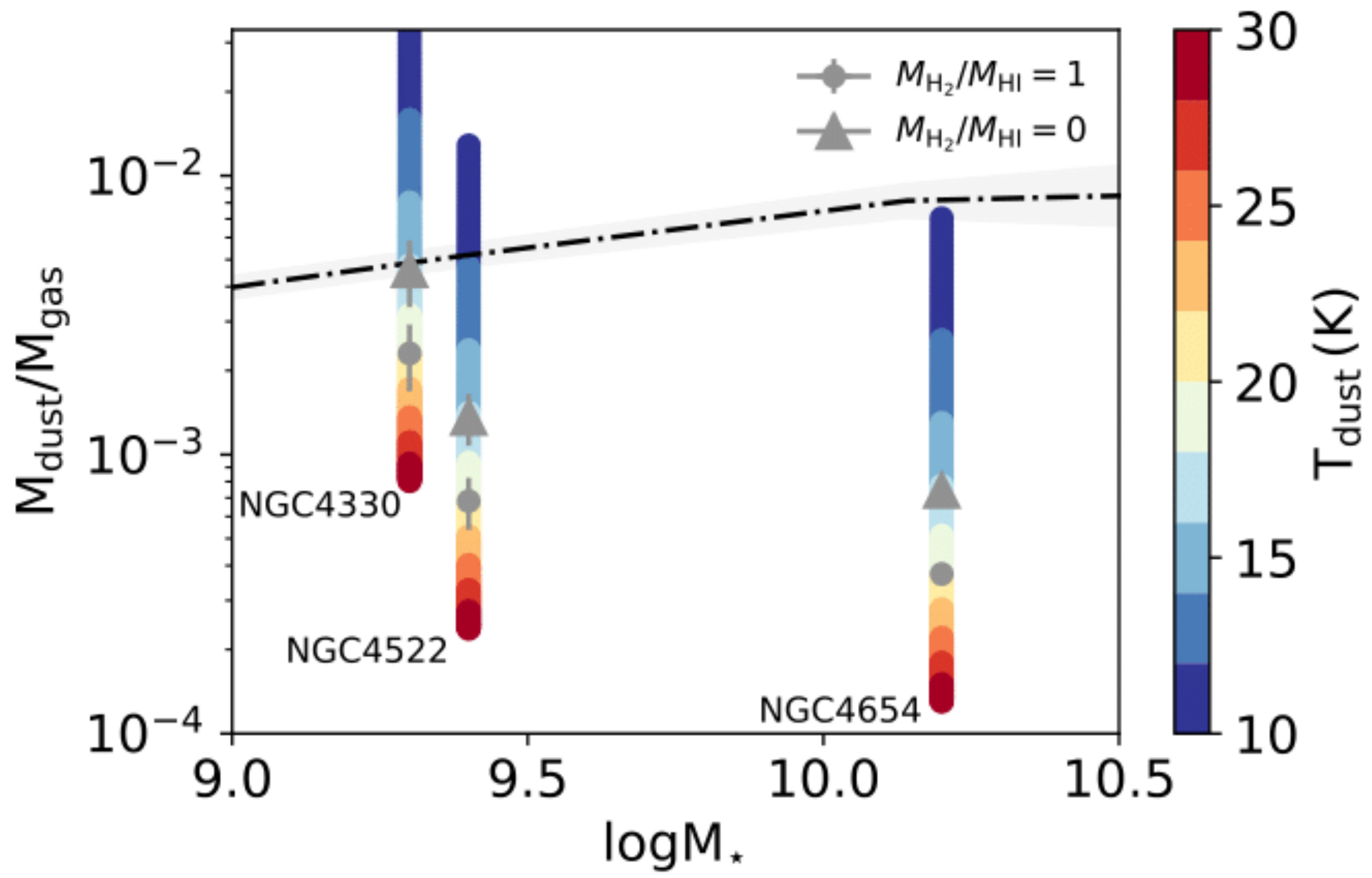}

\caption{
The dot-dashed line shows the dust-to-gas mass ratio vs. galactic mass measured in Virgo galaxies by \citet{Cortese2016}.  { Credit: \citet{Longobardi2020b}. A\&A, 644, A161, 2020, reproduced with permission \copyright~ ESO.} 
} 
\label{dustvirgoratio}
\end{figure}

From this point of view, it seems likely that in rich clusters, dust particles might be present in isolated fragmentary features manifested by the ICL clumps in the peripheral cluster regions. In this regard, it is worth remembering that \citet{Zwicky1951} concluded that {\it ``{... fainter nebulae are relatively more and more frequent as the distance from the center of the cluster increases.} 
''} This fact, confirmed recently in X-ray data \citep{Walker2019} and SDSS MgII absorbers data~\citep{Anand2022}, may indicate that the tidally stripped fragments in the cluster periphery remain relatively isolated without high-velocity merging and disintegration and as such can also contain interstellar gas along with dust; see the discussion by \citep{Mihos2004,Dolag2009}. Observations of the Virgo cluster also favor this conclusion. \citet[][]{Mihos2017} stress that tracers of extended \mbox{($\sim$$100$ kpc)} circumgalactic coronae can be found in galaxies of the cluster outskirts, while the signatures of tidal stripping are less pronounced in the core.  As mentioned earlier in Section \ref{latob}, \citet{Chelouche2007} and \citet{McGee2010} have also found indirect evidence of preferential residence of dust in peripheral regions of galactic clusters. Direct indications of this  can be found in measurements of dust mass concentration in the stripped tails of NGC 4330, NGC 4522, and NGC 4654 galaxies, that lie at 0.6, 0.9, and 0.9 Mpc of the Virgo cluster center \citep{Longobardi2020b}. The dust-to-gas mass ratio varies along the tails by two orders of magnitude from $\zeta_d\sim 10^{-4}-10^{-2}$ in NGC 4654 to $\sim$$2\times(10^{-4}-10^{-2})$ in NGC 4552, and $\sim$$8\times 10^{-4}-3\times 10^{-2}$ in NGC 4330; see Figure~\ref{dustvirgoratio}. Interestingly, the dust temperature in all cases anti-correlates with $\zeta_d$. It might indicate that dust heating is due to collisions with hot electrons (of the ICM or behind the tidally driven shock fronts), whereas a low dust-to-gas ratio is due to sputtering in collisions with hot ions--- the more intense the collisions, the higher the dust temperature and the lower the $\zeta_d$.      

Once tidally spread throughout the cluster, the dusty interstellar gas comes into contact with the ICM, and dust particles experience {damaging} impacts from hot ions, which induce sputtering from the dust surface; see, e.g., \citep{Draine1979,Jones1994,Dwek1996}. The characteristic sputtering time at $T>10^6$ K can be approximated for both graphite and silicate dust as $t_a(T)=a|da/dt|^{-1}\approx 10^{10}an^{-1}$ yr, $a$ is the dust grain radius (in cm), $n$ is the gas density; see Eq. 25.14 in \citep{Draine2011}. Thus, small dust grains with $a<0.01~\upmu$m that predominantly contribute to extinction deplete under typical ICM gas density $n\sim 10^{-3}$ cm$^{-3}$ in $\sim$$10$ Myr. 

Ram pressure $P_r=\rho_e v^2$, with $\rho_e$ being the density of external gas and $v$ being the velocity of a galaxy relative to external gas, is a well-known mechanism that is crucially important in shaping galaxies in clusters \citep{Gunn1972}. Head--tail ratio galaxies in clusters \citep{Miley1972,Miley1980,Cavaliere2013} are amongst the brightest examples; for later developments, see \citep{Garon2019}. 
More recently, ram pressure has been recognized to be also responsible for forming jellyfish galaxies; see the review in \citep{Muller2021}. Along with tidal stripping, ram pressure is a universally acting mechanism in mass exchange between the galaxies' ISM and the ICM; examples can be found, e.g., in~\citep{Boselli2006,Wezgowiec2012,Boselli2016,Moretti2021}. A very picturesque example has been obtained by \citet[][]{Boselli2018a} for several galaxies in the Virgo cluster; see also the discussion in the review \citep{Boselli2021}, two of them with stripped tails can be clearly seen in their Figure~9 \citep{Boselli2021}.
Ram pressure action manifests clearly in a $\sim$$150$ kpc long cometary tail to the right of the NGC 4569 galaxy along with the obviously undisturbed stellar component. Long HI tails $\simgt 100$ kpc are also seen in the galaxies NGC 4388 \citep{Oosterloo2005} and NGC 4424 in the Virgo cluster; a lower limit {of the tail  length} is because of a limited flux sensitivity~\citep{Boselli2018b}. An example of an exceptionally long HI tail in the Virgo---500 kpc driven by ram pressure has been demonstrated by \citet{Koopmann2008}, and another of {length} 250 kpc {has been} found recently in the outskirts of the Abell 1367 cluster \citep{Scott2022}. In general, a 100 kpc length of ram pressure-driven tail does not seem {to be} an overestimate, both for H$\alpha$ and HI tails \citep{Boselli2016,Boselli2021,Jachym2019}. Interestingly, a 100 kpc length scale also appears in the analysis of mutual distributions of MgII SDSS absorbers and neighbor galaxies in DESI clusters \citep{Anand2022}, indicating that in both cases, ram pressure is the most likely source for these structures. 

The lifetime of these structures can be roughly estimated as $\simlt$ 300 Myr assuming their velocities to be of $\simgt$300 km s$^{-1}$. However, note that for H$\alpha$ tails with an estimated density of $10^{-2}$ cm$^{-3},$ the recombination time is of $\sim$$10$ Myr \citep{Boselli2021}, suggesting that they are maintained for longer times by either of the two possible sources: young {in situ} born stars~\citep{Merluzzi2013,Steinhauser2016,Gullieuszik2020}---the possibility supported by direct observations of molecular gas in ram pressure-stripped tails \citep{Jachym2019} or by shock waves in the interface layer between the tail and the ICM (discussion and observational evidence are given in \citep{Merluzzi2013,Poggianti2017,Owers2012}) or by thermal conduction (see discussion in Section \ref{shockw}). It is worth noting in this regard that the ionizing background photons with $J_{\nu_0}\sim 10^{-21}$--$10^{-22}$ erg cm$^{-2}$ s$^{-1}$ Hz$^{-1}$ sr$^{-1}$ \citep{Haardt1996,Faucher2009} can support a layer of HII gas with the emission measure of \mbox{$EM\sim$ 1--10 pc cm$^{-6}$ \citep{Alnajm2016}}, and as such can support H$\alpha$ tails with density $n\sim 10^{-2}$ cm$^{-3}$ of thickness $\sim$10--30 kpc. Assuming the dust-to-gas ratio $\zeta_d\sim 0.1$ proportional to the ICM metallicity $Z\sim 0.3Z_\odot$, one can estimate the characteristic extinction of a tail with radius $R\sim 1$ kpc and density \mbox{$\sim$$10^{-2}$ cm$^{-3}$} \citep{Boselli2021} of the order $A_v\sim 0.003$. It decays as $A_v\propto R^{-1}$ when the tail mixes with the ambient ICM and falls down to $A_v\sim 0.001$ after mixing with ambient gas of \mbox{$\sim$$10^{-3}$ cm$^{-3}$}. In the prototypical Virgo cluster, a typical mass loss rate of tails is estimated as 1--10~$\msun$ yr$^{-1}$ with a characteristic lifetime widely varying from a few Myr to a few Gyr depending on their initial mass \citep{Taylor2020}. It is obvious that  tails with lifetimes shorter than 30--100 Myr lose their dust after being dissolved in the ICM and cannot contribute considerably to extinction or IR emission. On the other hand, long-lived tails are too rare to have a high covering factor and  cannot contribute much to extinction. \citet{Vollmer2007} argue that the observed deficiency of HI in spiral galaxies in the Virgo cluster implies a total mass loss $\sim$$10^9~\msun$ per galaxy under ram pressure stripping. Assuming the DtG mass ratio $\zeta_d\sim 10^{-2}$, this converges to the total dust mass added by spirals to the ICM gas in the Virgo $M_d\simlt 10^9~\msun$ for $\sim$100 spiral galaxies in the Virgo. The DtG mass ratio can potentially be $\zeta_d\simlt 5\times 10^{-5}$ when all the gas lost from spirals is diluted with the cluster gas, which eventually results in the extinction $A_v\simlt 0.01$. 

%
%

The question remains though as to how fast the stripped gas comes into thermal contact with the hot ICM. The efficiency of thermal contact depends on the surface $S_m$ of a given mass of stripped gas. In the beginning, $S_m$ is {likely} to be close to its minimum, while at later dynamical stages, hydrodynamical instabilities disturb the flow to bring it to a {more} irregular, turbulent state. In the simplest case of a cloud being accelerated or dragged in an ambient gas flow, the cloud becomes stretched opposite to its velocity with an increased surface area $S_m\sim 2\pi R_2\chi^{1/2}(t/t_{cc})$ \citep{Gronke2020}. Hydrodynamic instabilities stimulate a cascade of random motions from the larger scales downward; see the discussion in \citep{Klein1990,Klein1994,Elmegreen2004,Scalo2004,Evoli2011,Rennehan2019}. In case of tidal stripping, the energy cascade  begins from galactic scales $R_g\sim 30$ kpc and takes time of order $t_{cas}\sim R_g/v_t\sim 0.3$ Gyr, with $v_t\sim 100$ km s$^{-1}$ being the relative velocity of the galaxies involved into tidal interaction. One might think that this cascading time $t_{cas}$ limits the lifetime of dust in the ICM. However, the presence of a  magnetic field is commonly thought to change the situation.  

\subsubsection{ICM Magnetic Field}
When injected into the ICM from galaxies by tidal streams and/or by winds, dust is immersed in a hostile environment of hot plasma. Being coated by a cold/warm gas expelled from galaxies in the form of clouds or isolated flows, the dust is shielded from immediate destruction. Thermal conductivity tends to destroy this shield, and the destruction can be enhanced by a mutual action of hydrodynamical instabilities that stretch clouds and increase their surface area. The classical Spitzer conductivity $\kappa_{\rm Sp}=5\times 10^{-7}T^{5/2}$ erg s$^{-1}$ cm$^{-1}$ K$^{-1}$ ensures the hot gas to penetrate the cold layer of $L$ thick in $\tau_L\sim 1.5k_{\rm B}nL^2/\kappa_{\rm Sp}\sim 300n$ Myr for $T=10^7$ K, for $n=10^{-3}$ cm$^{-3}$ it is only $\tau_L\sim 0.3$ Myr. However, this conclusion is not {a firm one}. 

One of the most important barriers preventing the thermal evaporation of cold cloud in the hot ICM is the magnetic field, which can strongly inhibit thermal conductivity $\kappa_B\sim (\lambda_L/\lambda_C)^2\kappa_{\rm Sp}\sim 10^{-12}T_6^{-4}B_{1~\mu\rm G}^{-2}n^2\kappa_{\rm Sp}$ \citep{Rosseland1958}, where $\lambda_L$ is the electron Larmor radius, $\lambda_C$ is the electron Coulomb free path, and $\kappa_{\rm Sp}=5\times 10^{-7}T^{5/2}$ erg s$^{-1}$ cm$^{-1}$ K$^{-1}$ is the Spitzer conductivity; for more recent discussions, see \citep{Asai2004}.  
 
The presence of a rather strong magnetic field of $\sim$$0.1-1~\mu$G is confirmed in the observations of several groups. A lower limit on the volume averaged $B\geq 0.4-0.6\mu$G from inverse Compton emission is estimated for the A3667 cluster by \citet{Fusco2001}. \citet{Johnston2004} reports a tangled magnetic field of this order in the same cluster with the scale of $\sim$$100$ kpc in the central cluster region. \citet{Bonafede2010} infer from the rotation measure (RM) in the Coma intracluster space plasma in its center $\langle B\rangle\sim 4~\mu$G, assuming a Kolmogorov turbulence spectrum with the lower scale $\sim$$2$ kpc. Similar estimates from measurements of RM are obtained by \citet{Bohringen2016}. Quite recently, \citet[][]{XuandHan2021} and references therein found evidence of a $\sim$$4~\upmu$G magnetic field in the intracluster medium in clusters at $z>0.9$. 

However, in a turbulent magnetized medium, the reduction of the thermal conductivity by magnetic field seems to be less strong than in a regular magnetic field. \citet{Narayan2001} demonstrated that in a highly multi-scale turbulent medium, the conductivity is smaller than the classic value only by a factor of a few. Their consideration, though, is very simplified in the sense that the turbulence is assumed to be highly developed and steady state, whereas when a cold cloud is expelled (or stripped) from a galaxy into the ICM, it becomes surrounded by a hot gas with a local characteristics of turbulence, which is not necessarily highly developed and does not extend down to the dissipative scales; see, e.g., the discussion in \citep{Johnston2004}. In such conditions, the cloud can instead be draped by a regular field as predicted by \citet{Lyutikov2006} and developed further in, e.g., \citep{Asai2007} and references therein, while \citet{Sparre2020} described recently the draping of clouds by a random magnetic field. Moderate turbulence  with the velocity amplitude of $\sim$$100-300$ km s$^{-1}$ on scales $\sim$30--60 kpc in the Perseus cluster have been confirmed by {\it {Hitomi} 
} \citep{Hitomi2016}. Similar scales of turbulent motions 50--100 kpc in galaxy clusters have been also inferred from the measuring of X-ray brightness fluctuations \citep{Fujita2020}. These scales are readily seen to be in excess of typical sizes of dense clouds expelled from galaxies. 
    
\section{Conclusions}
The problem of the presence of dust in the intracluster medium is currently of considerable  interest, {despite}  a long history since the issue was first brought to attention by  Zwicky more than 70 years ago. Much theoretical (including numerical simulations) and observational efforts have been made during the recent 30--40 years, which have not only enriched  our understanding of the problem but allowed to {post} more focused questions to be undertaken for further understanding. Therefore, we can draw here only tentative conclusions. 
\begin{itemize}
\item Continuous mass and dust exchange between the ICM and the ISM of cluster galaxies is supported by galactic winds driven by shock waves from supernovae, stellar radiation, and radiation from active galactic nuclei, by cosmic rays, by gas stripped of galaxies in tidal interactions and ram pressure from the ICM. This exchange appears to be sufficiently powerful and  provides a channel for dust to fill the intracluster space. However, during the transport from galaxies to the ICM, dust experiences destructive action from sputtering in collisions with hot ions. 
\item {Dust particles can survive thermal destruction, at least partly, because during the transport, they are shielded in denser clouds or streams.} When the gas density in the interface between the cold gas in clouds and streams and the hot ICM gas is larger than $\sim$$0.3$ cm$^{-3}$, external heat flux can be radiated within a relatively thin layer and can stimulate local cooling of the ICM, protecting {the} confined dust. On the other hand, magnetic field on scales $\simlt$1--2 kpc, which is comparable or larger than clouds and tidal streams spatial scales, can suppress thermal evaporation.  
\item The role of the magnetic field in the protecting of clouds' thermal evaporation, as well as of hydrodynamical instabilities in the disintegration of them during their motion in the ICM is not entirely clear, and it needs to be further studied in numerical experiments.    
\item Observational measurements of the dust mass fraction in the ICM {are} rather challenging: in absorption, observational selection can be a reason because of a likely patchy distribution of dust throughout the intracluster space with a small covering factor, whereas the surface brightness of dust emission can be too weak for discriminating it from the CIB as the dust temperature is a rather low $T_d\sim 20$ K. In addition, contaminations from Galactic cirrus can critically disguise a weak dust emission. The spatial correlation of a dust-like FIR thermal emission with tSZ effect in clusters seems to be a promising, though observationally challenging, indicator of dust survival in there. 
\item Overall, current observational constraints indicate rather wide ranging values for the amount of  dust in the ICM, with extinction varying from $A_v\simlt 10^{-3}$ to $A_v\sim 0.2$. When direct measurements of the dust-to-gas mass fraction are possible, they imply $\zeta_d=10^{-5}$--$10^{-3}$, although this estimate remains 
beset by observational uncertainties and sensitivity limitations. {Among these, one of the most serious is connected to the dust type, as available data do not seem sufficient to distinguish between different~types. }
\end{itemize}
\vspace{6pt} 


\authorcontributions { The authors have contributed equally to this work. } 

\funding {The work of YS is done under partial support from the project ``New Scientific Groups LPI'' 41-2020. EV is thankful to the Russian Scientific Foundation (grant 19-72-20089).} 


\informedconsent { Not applicable. }


\dataavailability { Not applicable. }

\acknowledgments{This work has made use of NASA’s Astrophysics Data System Bibliographic Service. } 

\conflictsofinterest {{The authors declare no conflict of interest.}}

\abbreviations{Abbreviations}{
The following abbreviations are used in this manuscript: 

ICD---intracluster dust; ICM---intracluster medium; DtG---dust-to-gas; {\it IRAS}---InfraRed Astronomical Sattelite; {\it COBE}---Cosmic Background Explorer; ISO---Infrared Space Observatory; SDSS---Sloan Digital Sky Survey; DESI---Dark Energy Spectroscopic Instrument; max-BCG catalogue is a catalogue of galaxy clusters based on the SDSS maxBCG selection algorithm; tSZ---thermal Sunyaev--Zeldovich effect; SED---spectral energy distribution; ISM---interstellar medium, CGM---circumgalactic medium; TW---thermal wave; PAH---polycyclic aromatic hydrocarbon; CR---cosmic rays; SFR---star formation rate; ICL---intracluster light; LBT---LArge Binocular Telescope; LPI---Lebedev Physical Institute. ULIRG---Ultraluminous InfraRed Galaxy. }

\end{paracol}
\printendnotes[custom]
\reftitle{{References}}


\end{document}